\documentclass{aa}
\usepackage{amssym}
\input epsf
\begin{document}
\thesaurus{08                   %A&A section 8: diffuse matter in space
             (02.19.2;          %scattering
              09.09.1 30 Dor;   %ISM: individual: 30 Dor
              09.04.1;          %ISM: dust, extinction   
              11.09.1 LMC;      %galaxies: individual: LMC
              11.09.4)}         %galaxies: ISM
\title{Extended Red Emission (ERE) detected in the 30 Doradus nebula}
\author{S. Darbon, J.-M. Perrin and J.-P. Sivan}
\thanks{Based on observations made at the European Southern Observatory (ESO),
La Silla (Chile)}
\offprints{J.-P. Sivan}
\institute{Observatoire de Haute Provence du CNRS, F-04870 Saint Michel 
l'Observatoire, France}
\date{Received xxx ; accepted yyy}
\maketitle
\begin{abstract}
We present low-dispersion visible spectra of two regions of the 30 Doradus 
nebula. When corrected for atomic continuum emission and divided 
by the spectrum of the exciting and illuminating stars, one of the two
spectra clearly shows Extended Red Emission (ERE) superimposed on the
scattering component. This ERE band peaks around 7270 \AA~ and is 1140 \AA~ 
width. HAC grains are found to explain the observed spectra in terms of
scattering and luminescence.
\end{abstract}
\section{Introduction}
\noindent
Spectrophotometric studies have revealed the presence of a wide emission band
in the red part of the continuous spectrum of several types of 
galactic objects : reflection nebulae
(Witt \& Boroson 1990), planetary nebulae (Furton \& Witt 1992), HII regions
(Sivan \& Perrin 1993) and high latitude cirrus (Guhathakurta \& Tyson
1989); this band, called Extended Red Emission
(ERE), consists of a broad emission
bump about 1000 \AA~ wide (FWHM), centered between 6000 and 8000 \AA.
In order to explain this emission, two hypotheses are generally
invoked : Poincar\'e fluorescence of isolated Polycyclic Aromatic 
Hydrocarbons (PAH) 
molecules (d'Hendecourt et al. 1986, L\'eger et al. 1988) or fluorescence 
from solid state materials such as filmy Quenched
Carbonaceous Composite (f-QCC) (see e.g. Sakata et al. 1992) or 
Hydrogenated Amorphous Carbon (HAC) (see e.g. Watanabe et al. 1982, 
Furton \& Witt 1993).

In order to better characterize the mechanisms responsible for ERE, we have thought
it useful to observe objects in which the physical and 
chemical conditions are distinct from those encountered in galatic nebulae. 
This has led us to observe the dusty halo of the galaxy M82 in the
spectrum of which we have detected an ERE band (Perrin et al. 1995). Also, we
have observed HII regions in the Magellanic Clouds because 
the interstellar medium of these galaxies shows a lower metallicity than
the interstellar medium of our Galaxy (Russell \& Dopita 1990). Moreover, their
proximity make them well suited for detailed studies. In this paper, we 
present the results obtained on the well-studied
starburst nebula 30 Doradus in the Large Magellanic Cloud (LMC). This 
supergiant HII region has a bright core of 
200 pc in diameter with filaments outlying over 2 kpc. Its morphology and 
dynamics are very peculiar (Elliott et al. 1977, Cox \& Deharveng 1983). 
Its stellar content is uncommon ; in particular, the R136 cluster 
is rich in WR and early type O stars and releases almost 50\% of the ionizing 
flux in the nebula (Walborn 1984, Melnick 1985).

The observations and data reductions are described in the next section. Section
3 presents the main spectrophotometric results we have interpreted in terms
of dust scattering and fluorescence.

\section{Observations and data reduction}
%**********************
\begin{figure*}
\epsfbox{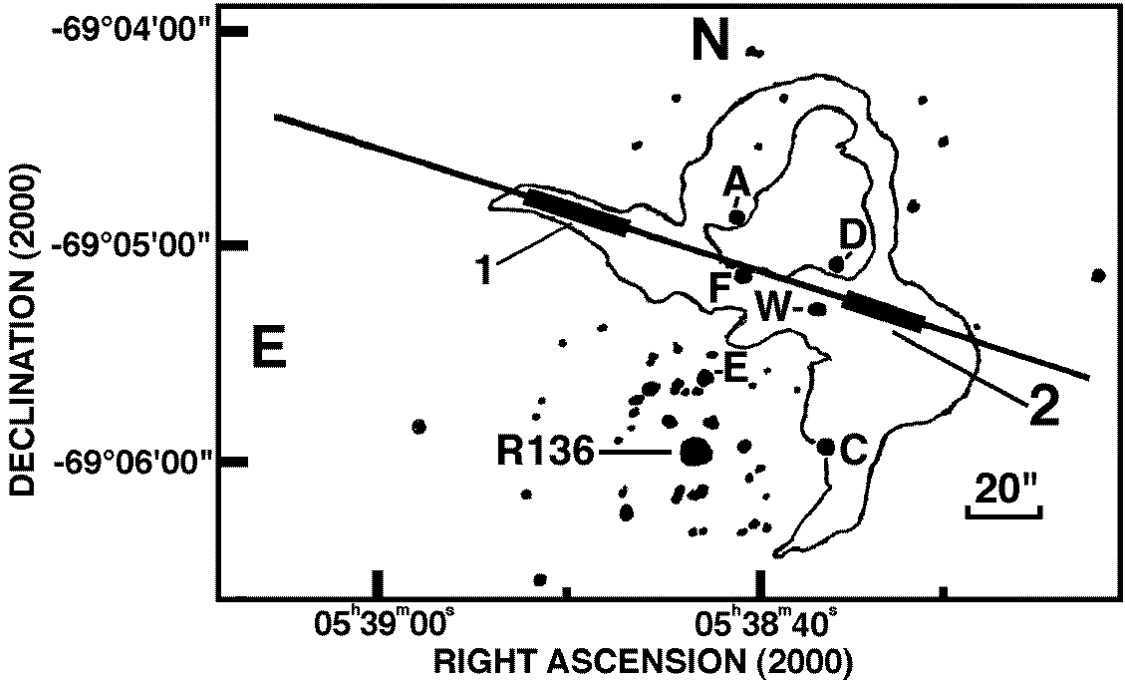}
\caption{Spectrophotometric observations of the 30 Doradus HII region :
schematic representation of the position of the spectrograph slit. Regions 1
and 2 for which one-dimensionnal spectra were extracted are shown. The ionizing
and illuminating sources are indicated : R136, D, C, W, F, A and E, as 
labelled by Melnick (1983)}
\label{fig1}
\end{figure*}
%**********************
%**********************
\begin{figure*}
\epsfbox{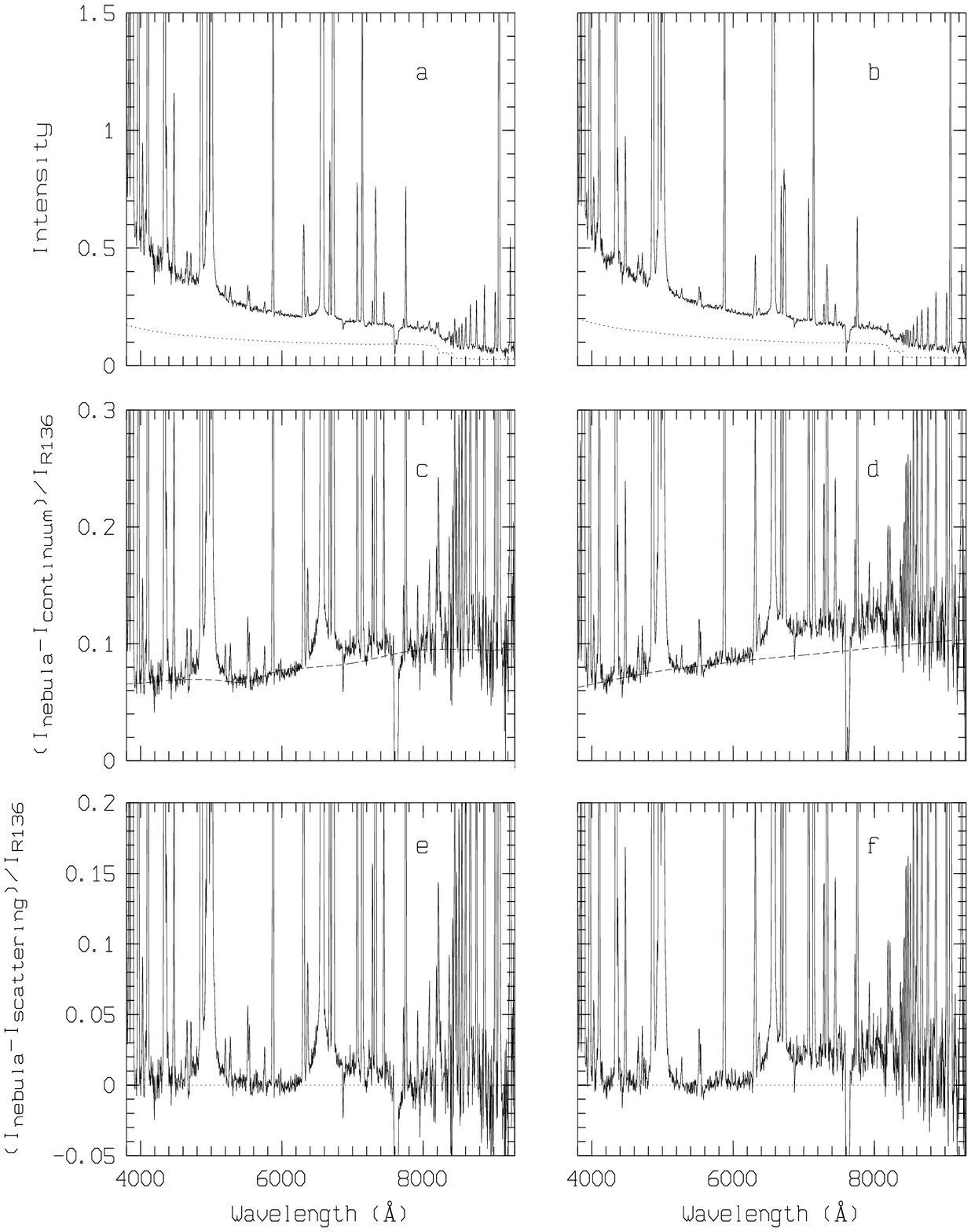}
\caption[]{(a) and (b) Spectra of, respectively, region 1 and region 2 of 30
Doradus ; the theoretical atomic spectrum is superimposed in dotted line
(Intensity unit : (a) 10$^{-13}$erg cm$^{-2}$ \AA$^{-1}$ s$^{-1}$ arsec$^{-2}$, (b)
10$^{-12}$erg cm$^{-2}$ \AA$^{-1}$ s$^{-1}$ arsec$^{-2}$).
(c) and (d) Division of the nebula spectrum corrected for the atomic 
continuous emission by the spectrum of the exciting stars ; theoretical 
scattering spectra are superimposed in dashed line.
(e) and (f) Subtraction from the observed scattering spectra of theoretical 
scattering spectrum. The strong absorption A band of O$_2$
is apparent in all spectra around 7620\AA.}
\label{fig2}
\end{figure*}
%*********************
\noindent
Spectra of the 30 Doradus nebula in the LMC were obtained in November
1994 using the Boller \& Chivens long slit spectrograph equiped with a
Fairchild CCD (FA 2048L) and mounted at the Cassegrain focus
of the European Southern Observatory  1.52 m telescope in La Silla. The 
spectral domain of interest [3800 \AA,9300 \AA] was covered with a mean 
dispersion of 250 \AA~mm$^{-1}$. The usable slit length on the sky was 226" 
(276 pixels on the detector), the slit width was set to 3", which corresponds 
to a spectral resolution of 13.7 \AA.
A single slit position (position angle of 77$^\mathrm{o}$) was observed on the
nebula. It was centered 57" north of
R136 (Fig. \ref{fig1}). We made a number of short exposures, alternatively on 
the nebula and on the surrounding sky background located 93' 
north-east from the position on the nebula. 
As discussed previously (Perrin \& Sivan 1992), this procedure avoids the
saturation of the CCD with bright nebular emission lines and allows a fine
monitoring of the continuous spectrum of the night-sky background. The
following sequence was taken up: 1 exposure of 120 s on the sky background + 1 
exposure of 240 s on the nebula + another exposure of 120 s on 
the sky background.
A total exposure time of 960 s was obtained on the nebula. The exciting and
illuminating cluster R136 and the surrounding brightest stars D, C, W, F, A 
and E (which are thought to contribute to the illumination of the nebula),
as quoted in Fig. \ref{fig1}, were observed near the meridian 
in order to compare their spectral energy distribution with 
that of the nebula. The standard star LTT377 
(Hamuy et al. 1992) was observed for flux calibration.

Data reduction was conducted
at Observatoire de Haute Provence using MIDAS software on a DEC-ALPHA
computer. After co-adding the short exposure frames, offset subtracting and 
flat-fielding, the spectrum was spatially divided into two bins which correspond
to two different parts of the nebula (free of contamination by stellar spectra)
and one-dimensionnal spectra were extracted for these two regions. Region 1 is 
located 33" east and 67" north from R136 and its dimensions are 
28"$\times$3" ; region 2 is located 52" west and 38" north from R136 
and its dimensions are 22"$\times$3" (Fig. \ref{fig1}). Then, each 
one-dimensionnal spectrum was 
wavelength- and flux-calibrated. Atmospheric extinction correction was applied 
using the standard extinction curve of La Silla. Sky subtraction was done by 
removing from a nebular spectrum the corresponding sky background spectrum 
extracted by binning the same pixels along the spectrograph slit (in order to 
avoid any instrumental function effect). The spectrum of each exciting star 
was reduced as above except for the sky subtraction : we subtracted a mean 
sky background taken on each side of the star spectrum on the same frame.

In the visible range, the LMC interstellar reddening curve (Nandy et al. 1981,
Morgan \& Nandy 1982) 
does not differ from the standard curve (Pr\'evot et al. 1984, Howarth 1983,
Pei 1992). So, the spectra were dereddened using the LMC curve 
and values of the interstellar extinction were measured globally to take
account of the galactic and LMC interstellar extinction. We applied the
classical formula that gives the nebular intensity I$_\mathrm{o}(\lambda)$ 
relative to H$\beta$ corrected for interstellar extinction as a function of 
the observed intensity I($\lambda$) and the observed H$\beta$ intensity 
I(H$\beta$) :\\
\begin{equation}
 \mathrm{I_o(\lambda)/I_o(H\beta) = [I(\lambda)/I(H\beta)]10^{C(H\beta)f(\lambda)}}
\end{equation}
C(H$\beta$) and f($\lambda$) are respectively the reddening correction at
H$\beta$ and the reddening function. For the two regions on the nebula,
C(H$\beta$) was derived from the comparison of the observed
H$\delta$/H$\beta$, P9/H$\beta$ and P11/H$\beta$ ratios with the theoretical 
ratios computed by Brocklehurst (1971) under
case B conditions. We obtained C(H$\beta$)=0.58 for region 1 and
C(H$\beta$)=0.35 for region 2. 
We note that these values compare well with those found by Mathis et al. 
(1985). For R136, we used the value C(H$\beta$)=0.49 from Fitzpatrick \& Savage
(1984).

For each region, we have calculated the atomic continuum emission, 
normalized to 
H$\beta$, supposing a pure hydrogen nebula ; since N(He$^{+}$)/N(H$^{+}) 
\approx$ 0.082 (Rosa \& Mathis 1987), 
the He contribution can be neglected. The continuous atomic spectrum 
is the sum of the
two-photon decay emission and of the free-bound continuum radiation and it is a
function of the electron temperature and of the electron density. We
measured the electron temperature and the electron density in a classical
way, using respectively the [NII] and [SII] lines intensity ratios.
We found T$_e$=10250 K and n$_e$=800 cm$^{-3}$ for region 1 and T$_e$=11600 K 
and n$_e$=200 cm$^{-3}$ for region 2. The HI recombination continuous emission
and the HI two-photon continuum were calculated using the formulae given by
Brown \& Mathews (1970). To calculate the first component we used the HI
continuous emission coefficient obtained using a 
spline interpolation from the data
of Brown \& Mathews (1970). For the calculation of the second component we used
the effective recombination coefficient for populating the $2^{2}S$ level of HI
and the collisional transition rate for HI 2$^2$S, 2$^2$P, both interpolated for
the same temperature from the same data. The total continuum spectrum was
normalized to the intensity of the H$\beta$ line. Also, to
avoid artifact in the subtraction of the atomic continuum, we convolved the
theoretical spectrum with the instrumental function (i.e., a 13.7 \AA~window)
from 8199 to 8447 \AA, thus taking into account the instrumental blend of the
Paschen lines down to Pa18. The intensity of the Paschen lines was 
calculated relative to the intensity of the H$\beta$ line, using the
formula given by Goldwire (1968). 
Figures 2a and 2b show the dereddened spectra for respectively regions 1 and 2. 
The calculated atomic continuous spectra are superimposed in dotted line. 

To characterize the scattering and, eventually, 
the luminescence phenomena, we
have compared the color of the nebula spectra to 
that of the illuminating source
spectrum. For this purpose, we have divided the dereddened spectra of regions 
1 and 2, corrected for atomic continuum emission, by the sum of the dereddened 
spectra of R136 and of the surrounding brightest stars D, C, W, F and A. The 
resulting spectra are displayed in figures 2c and 2d. In what follows,
we shall refer to them as \textit{reduced spectra}. As shown by Hill et al.
(1993), stars in the close vicinity of R136 have very low nebular extinction
($E(B-V)<0.03$). As a result, the reddening between the illuminating
sources and regions 1 and 2 is negligeable and the reduced spectra 
in figures 2c
and 2d do reveal the scattering and luminescence properties of the nebular
dust.

\section{Results and discussion}
\noindent
These spectra appear to be basically red over the entire 4000-9000\AA~
wavelength range. 
This can be simply explained by a significant contribution to scattering from
dust particles whose size parameter is greater than 1. According to Lamy \&
Perrin (1997), radiation pressure acts more efficiently on submicronic grains
than gravitational attraction does. But this is not the case for micron-sized
grains. So, in the neighbourhood of early-type stars, in particular in the
region surrounding R136, submicronic grains are expelled, leading to a decrease
of the grain density and to an increase of the relative abundance of large
grains. This is in good agreement with the weak internal extinction observed by
Hill et al. (1993).

We have tried to model the scattering component in each reduced spectrum.
For this purpose, we have studied scattering of light by grains for a large
number of materials which are real materials of astrophysical interest :
silicates (olivine, andesite, augite, ...), metals (iron, nickel), iron
sulphides (pyrrhotite, troilite), organic materials (tholins, icetholin,
kerogen, poly-HCN), glassy carbon, amorphous carbons, graphite, a lot of 
HACs (whose properties depend on the conditions of synthesis), silicon 
carbides and meteoritic materials.
Calculations were conducted in the framework of single scattering (justified by
small values of optical depth in the observed regions) 
for several grain size distributions and several
scattering angle intervals. We retained only spectra that (i) fit the observed
ones at least in their bluest ($\lambda \lesssim 5500$\AA) and reddest 
($\lambda \gtrsim 8500$\AA) parts, and (ii) change only
smoothly when scattering parameters are slightly modified (i.e. stationnary
solutions).
Among all the calculated spectra, only those obtained with HAC grains proved to
fit satisfactorily the observed scattering components.

For region 1, the best fit (Fig. 2c) was obtained for HAC grains (i) which have
undergone a very low bombardment by ions and electrons during their formation, 
(ii) which have never been submitted to high temperature and (iii) for
which minimal adsorption of hydrogen atoms has occured. Figure 2e shows the 
observed reduced spectrum after  subtraction  of the theoretical component. 
The ERE band does not appear very clearly : if it exists, it should be 
extremely faint. The fit is obtained using the complex indexes of refraction
given by McKenzie et al. (1983), a maximum particle radius of 0.5 $\mu$m,
a value of 2.9 for the exponent $p$ of the size distribution ($\propto a^{-p}$
where $a$ is the radius of the particle) and a wide interval of scattering
angle ($\simeq$ [10$^\mathrm{o}$,170$^\mathrm{o}$]).
This value of $p$, which is lower than those
found in the diffuse interstellar medium ($p \simeq 3.5$), is in agreement with
the relative abundance of micron-sized grains.

For region 2, we have found that the best fit to the scattering
component was given (Fig. 2d) by a model using HAC grains submitted to high
ionic bombardment and high hydrogen excess during their formation. The fit is
obtained using the complex indexes of refraction given by Khare et al. (1987), 
a $p$ value of 1.5,  a maximum particle radius of 1.5 $\mu$m and
a  small scattering angle range  ($\simeq$ [75$^\mathrm{o}$,105$^\mathrm{o}$]).
In this case, an ERE band is unambiguously detected (Fig. 2f). 
It looks similar to those found in galactic HII regions
(Perrin \& Sivan 1992, Sivan \& Perrin 1993), with a peak wavelength of 7270 
\AA~ and a width of 1140 \AA~ (FWHM). The ratio of integrated ERE intensity in
the 6000-9000 \AA~ range to the scattered light intensity is $0.20\pm
0.05$. This value is in the average of those found in reflection nebulae 
(Witt \& Boroson 1990).

The location of region 1 relative to R136 and the other luminous stars
(Fig. \ref{fig1}) is fundamentally unknown but is not incompatible with the
wide range of scattering angle needed to fit the observed scattering
component: we can suppose that most of the stars contribute to the 
illumination of this region. On the contrary, the narrow angular range
used for region 2 suggests  a contribution from  a smaller number
of illuminating stars.
This range being centered at 90$^\mathrm{o}$, the observed color
of the scattered light appears redder than it would be if the
scattering angular range was wider (Witt 1985, Perrin \& Lamy 1989).
Consequently, the intrinsic values of $p$  and of the
maximum particle radius should be respectively greater and lower 
than those  of our fit. It should be noted that
color effects can also be induced by surface roughness of grains
(Perrin \& Lamy 1989).

The properties of HACs, in particular their optical properties, depend on the
conditions of preparation (see e.g. Watanabe et al. 1982, McKenzie et al. 1983,
 Khare et al. 1987) and on the radiative, ionic, atomic
and/or molecular processings that follow their formation 
(see e.g. Furton \& Witt 1993). When
submitted to UV radiations, these materials exhibit luminescence phenomena,
whose intensity is releated to the UV illumination conditions and to the
efficiency of rehydrogenation of the grains (Duley et al. 1997).

Accounting for ERE by luminescence from HAC grains, our results appear to be 
self-consistent on the basis of laboratory results, in the sense that 
ERE is found to be relatively bright
in a region where the scattering is explained by
highly hydrogenated HAC grains, and, on the contrary, 
ERE is absent (or barely detected) in a region where the 
scattering is accounted for by poorly hydrogenated HAC grains.

Also, our results suggest that astrophysical conditions should significantly
differ from one region to the other. This is probably the case,
since hydrogen molecules are present
in region 1 and absent in region 2, according to the H$_2$ emission
map of Poglitsch et al. (1995).
ERE is found in a region where H$_2$ is absent and inversely.
This result agrees well with the finding of Field et al. (1994) and Lemaire et 
al. (1996) who did not observe any clear correlation between 
ERE intensity and H$_2$ emission in NGC 2023 and NGC 7023.

For the two observed regions, HAC grains are invoked to explain the
scattering and luminescence components. We think that the presence of such carbonaceous grains in
the LMC interstellar medium is supported by the two following facts :\\
(i) In the LMC, the stellar carbon abundance and, in particular, the stellar 
C/O abundance ratio are entirely normal but the HII regions appear to be
overdeficient in carbon and have a C/O ratio lower than that observed in the
Galaxy (Russel \& Dopita 1992). This suggests 
that a substantial fraction of the interstellar
carbon might be in the grains, in agreement with the good spatial correlation
observed between the emission of CO, C$^+$ and the dust thermal emission in the
far infrared (Poglitsch et al. 1995).\\
(ii) A very weak ultraviolet extinction bump is observed in the 30 Doradus
region (Fitzpatrick 1988). As shown by Colangeli et al. (1993), 
HAC grains do not exhibit an extinction feature at 2175 \AA.\\

\section{Conclusion}
\noindent
We have observed the continuous spectrum of two regions in the core of the 30 
Doradus nebula. ERE is unambiguously revealed for one of the two regions and is
at the limit of detection for the other region. This is the first detection of 
ERE in an extragalactic HII region.

The differences in the observed spectra cannot be accounted for by variation
in the internal extinction across the nebula : it has been shown by Hill et al.
(1993) that extinction is almost negligeable in the central part of the 30
Doradus nebula, a finding which is in good agreement with the morphology and
the dynamics of the nebula.

HAC grains can explain the observed spectra in 
terms of scattering and luminescence : highly hydrogenated HAC grains and 
poorly hydrogenated HAC grains are invoked respectively for the region where 
ERE is clearly visible and the region where ERE is not obviously seen. This 
difference in grains properties is in good agreement with the difference in the 
astrophysical conditions prevailing in the two regions.

%************************************
\it Acknowledgement. \rm We wish to thank the referee, Pr. A. Witt, for helpful
comments.
%***********************************

\end{document}